\documentclass{aa}  

\usepackage{graphicx}
\usepackage{multirow}
\usepackage{txfonts}
\begin{document}

   \title{Search for pairs and groups in the 2006 Geminid meteor shower}

   \author{P. Koten \inst{1}
		  \and
		  D. \v{C}apek \inst{1}
		  \and 
          P. Spurn\'{y} \inst{1}
          \and
          R. \v{S}tork \inst{1}
          \and
          V. Voj\'{a}\v{c}ek \inst{1}
          \and
          J. Bedn\'{a}\v{r} \inst{2}
          }

   \institute{Astronomical Institute, CAS, Fri\v{c}ova 298, 25165 Ond\v{r}ejov, Czech Republic\\
              \email{pavel.koten@asu.cas.cz}
              \and
              {Faculty of Electrical Engineering, Czech Technical University, Prague, Czech Republic}
              }

   \date{Received dd-mm-yyyy; accepted dd-mm-yyyy}

 
  \abstract
   {The question of the existence of pairs and groups among the meteor showers is opened for a long time. The double station video observation of the 2006 Geminid meteor shower, one of the most active annual showers, is used for the search of such events.}
   {The goal of the paper is to determine whether the observed pairs of Geminid meteors are real events or cases of random coincidence.}
   {The atmospheric trajectories of the observed meteors, photometric masses, and both time and spatial distances of meteoroids in the atmosphere were determined using a double station video observation. Time gaps among them were analysed statistically. The Monte Carlo simulation was used for the determination of the probability of random pairings.}
   {Higher than expected number of candidates for pairs was found among 2006 Geminids. Evaluation of Poisson distribution shows that a significant fraction of them may be real cases. However, the Monte Carlo simulation did not confirm this result and provided a different view. Analysis of geometrical positions of candidate pairs also did not support the presence of real pairs and groups. Although we cannot exclude that some of them may be physically connected pairs, all the observed cases can be explained as the coincidental appearance of unrelated meteors.}
   {}


   \titlerunning{Search for the Geminid pairs}
   \authorrunning{P. Koten et al.}

   \maketitle
%

\section{Introduction}

Visual and telescopic observers repeatedly reported that the meteors appear in pairs or even groups. As these observational methods are rather subjective, more reliable results can be provided by instrumental observations. Earlier studies of the major meteor showers and sporadic meteors usually lead to contradictory results. \cite{Porubcan1968} investigated radar records of 6 meteor showers -- including Geminids -- using several statistical methods and found an excellent agreement with a random time distribution. A comprehensive overview of earlier studies is provided by \cite{Porubcan2002}. Their paper also analysed recent Leonid storms observed in 1966, 1969, and 1999 as well as April Lyrid activity observed by radar in 1982. The authors found positive results for younger streams. In the case of older showers, no grouping of meteoroids above the random level was found. Similarly, clustering analysis of the arrival times of 1999 Leonids did not show any clustering on the time scales of video frame rate \citep{Gural2000}. The authors conclude that meteoroids do not tend to fragment in the vicinity of the Earth. \citet{Ofek1999} found a very small excess of clustering on 5-seconds intervals when analysing 1998 Leonid visual and video observations, but did not suppose the results as conclusive.

On the other hand, several papers reported observations of the meteor clusters. \citet{Hapgood1981} analysed double station television observation of three Perseid meteors arriving within 1.3~seconds on August~12, 1977, and interpreted this event as a result of the fragmentation that occurred at least 1700~km from the Earth surface. Five nearly simultaneous meteors were observed within only 0.1~s on October~18, 1985 \citep{Piers1993}. Their radiants were not consistent with any major meteor shower. Moreover, each main object seemed to be composed of several smaller fragments. Huge clusters were observed several times during recent Leonid meteor storms at the turn of the century. \citet{Kinoshita1999} reported 100-150 Leonid meteors observed within 2~s during the 1997 return, and \citet{Watanabe2002} at least 15 meteors within 4~seconds in 2002. Both events were only single-station observations. In another paper \citet{Watanabe2003} provided results of 38 double station Leonids observed again in 2002 which occurred within 2~s. They analysed all three Leonid events, assumed them as the meteor clusters, and discussed possible scenarios of their origin.

Another interesting event occurred above the Czech Republic on September~9, 2016 \citep{Koten2017}. It consisted of a bright fireball and 8 faint meteors all of them flying on parallel trajectories. All the meteors were recorded within 1.5~second. The double station observation provided their atmospheric trajectories -- the spatial separation among them was between 14 and 105 km. The small meteoroids were supposed to be fragments separated from the main body no earlier than 2 or 3 days before entering the atmosphere at distances smaller than 0.08~AU. The meteor cluster belonged to the September $\epsilon$ Perseid meteor shower.

Here we investigate the possibility of the existence of the meteor pairs and groups within the Geminid meteor shower. This is one of the most intensive annual showers (004 GEM\footnote{IAU Meteor Data Center code for Geminids}). It is usually active in the first half of December with a peak around December 14. In recent years the zenith hourly rate is reaching values of about 120-140 meteors. \citet{Plavec1950} was the first, who noted that the observability of the Geminid meteor shower from the Earth is limited to few centuries. The first observation of the Geminid meteor was found to be recorded in 1077 in the Chinese chronicles \citep{Hasegawa1999}. The age of the stream is no longer that few thousand years. Its activity is increasing in the last few decades \citep{Ryabova2018}. \citet{Jenniskens2006} suggests that it will reach its maximum around 2050. The model of \citet{Ryabova2016} implies the cometary origin of the meteor stream. On the other hand, the observations of the meteors in the atmosphere did not support such origin. For example, fireballs reported by \citet{Spurny1993} exclusively belonged to the fireball group I called ''asteroidal meteors''.

Asteroid (3200) Phaeton is supposed to be a parent body of the meteor shower as was firstly suggested by \citet{Whipple1983}. It is probably a dormant comet which already lost all the volatiles and does not replenish the meteoroid stream anymore. Only weak and very short perihelion activity was observed in 2009 and 2012 \citep{Li2013} but this cannot contribute to the current activity of the meteor shower \citep{Ryabova2018b}.

Traditionally, statistical methods were used for the analyses of possible groupings among the meteors \citep{Porubcan1968}. They rely mainly on the Poisson distribution, which shows the number of expected random appearances of the meteor pairs or groups. Another way is a comparison of interval distribution between consecutive meteors compared with exponential distribution. Nevertheless, doubts were cast on those methods by \citet{Sampson2007} who found that there is a significant discrepancy between the Poisson statistics and numerical models. The Poisson distribution uses fixed time intervals what is not the case of typical meteor observations. The author found that the Poisson distribution underestimates the number of coincidental meteor groupings. He recommended further testing on the Perseid and Geminid meteor showers as they show consistent activity profiles. Therefore, we provide a comparison of both approaches to the solution of the problem based on the 2006 Geminid meteor shower observations. 

The paper is organized as follows: Section \ref{processing} describes observations, instrumentation, data processing, and the observed data on the meteor pair candidates, Section \ref{results} a comparison between statistical analyses of the pairs and Monte Carlo simulation as well as the geometrical properties of potential pairs, and Section \ref{conclusions} summarizes the results.

\section{Data acquisition, processing and overview}
\label{processing}

The double station video observations were carried out at the Ond\v{r}ejov Observatory since 1998. The analogue recording was switched to digital in 2008. Recently, the archive videotapes were digitalized and each meteor record was stored in the form of an AVI video sequence. When checking this process, some pairs were recognized. That initiated the idea to investigate whether the pairs are real events or rather random coincidences.

The S-VHS cameras connected with Mullard XX1332 image intensifiers and 50~mm lenses were used for the observations during the era of the analogue recording. As usual, the cameras were deployed at the Ond\v{r}ejov (14$^{\circ}$ 46$^{'}$ 48.5$^{''}$ E, 49$^{\circ}$ 54$^{'}$ 36.1$^{''}$ N, 524 m) and Kun\v{z}ak (15$^{\circ}$ 12$^{'}$ 1.1$^{''}$ E, 49$^{\circ}$ 6$^{'}$ 27.3$^{''}$ 
N, 650 m) stations. Distance between them is 92.5~km, the azimuth of the southern station is 340$^{\circ}$ \citep{Koten2004}. The data was stored on S-VHS videotapes, which were searched using automatic meteor detection software MetRec \citep{molau1999}. Meteor records were digitalized with a resolution of 25 images per second and 768 x 576 pixels. 

The shower membership was determined using MetRec. For the whole night, the time gap between consecutive Geminid meteors was computed. When this gap was shorter than 2 seconds, both meteors were listed as candidates for a pair. Such cases were measured manually using the MAIAMetPho software \citep{Koten2016} and their atmospheric trajectories and heliocentric orbits calculated using the method of \citet{Boro1990}. In some cases, one or both members of the possible pair were recorded only at one station. For the single station meteors, their atmospheric trajectories were computed under the assumption that the velocity is the same as the Geminid shower velocity. The meteor radiant was compared with Geminid nominal radiant. If the distance between them was smaller than $5^{\circ}$, the meteor was supposed as a Geminid meteor and thus a member of a pair candidate. Although the precision of the atmospheric trajectory determination in such cases is lower in comparison with the double station meteors, in this kind of study it is not so important. Having a higher distance, the meteor was not considered as Geminid and the pair was dismissed from the list. The average radiant  and velocity ($\alpha_{G} = 113^{\circ}, \delta_{G} = 33.3^{\circ}, v_{G} = 34.3~km/s$) based on the values posted for the Geminids on the IAU Meteor Data Center web page \footnote{https://www.ta3.sk/IAUC22DB/MDC2007/} were used as nominal radiant and velocity for the single station meteors.
  
\begin{figure}
  \resizebox{\hsize}{!}{\includegraphics{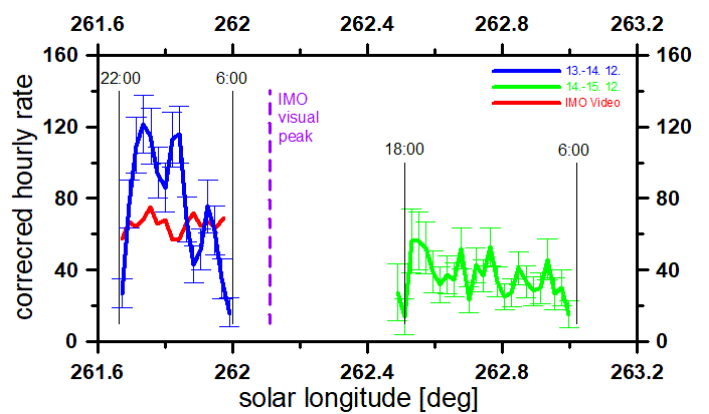}}
  \caption{The activity of the 2006 Geminid meteor shower determined using the video cameras from December 13th to December 15th. Thick dashed violet line represents the peak determined by IMO VMDB, thin black lines time UT. Red line shows the activity profile derived by the IMO Video Meteor Network.}
  \label{fig_gemchr}
\end{figure}

\subsection{Recorded data}
\label{data}

According to the IMO\footnote{\label{note1}https://www.imo.net/} Visual Meteor Database (VMDB), the maximum of 2006 Geminid meteor shower occurred in the morning hours of December~14. The peak was measured to occur at around 8:38~UT with ZHR$\sim$ 130 meteors. Visual inspection of the video records showed a higher number of potential meteor pairs during the night before the maximum. Therefore, this night as well as the night following the maximum were searched for the potential pairs. Conditions were cloudy at the beginning of the first night, thus the observation started later. Also, in the morning hours of December 14th, the conditions were not ideal. Nevertheless, during 8 hours of observations altogether 720 single and double station meteors were recorded. 524 of them were Geminid meteors, 251 recorded at both stations, 273 at one of them. Within the night following visual maximum, the observational conditions were better but the number of Geminid meteors already started to decline. Although the observational period was 50\% longer, the total number of 730 observed meteors was similar to the previous night and the number of Geminids decreased to 320.

\subsection{Geminid activity}

For the statistical analyses of the meteor pairs' existence, the shower activity is necessary to be known. \citet{Ryabova2018} showed that the activity of the Geminid meteor shower is increasing for the last 30 years. According to Figure 3 in this paper, the visual ZHR was between 130 and 140 in 2006.

We did not use those values because they are based on visual observations. The activity reported by visual observers is not directly comparable with the video rates. Both methods offer different sensitivity and field-of-view (FOV) what can cause the absolute numbers of the meteors to be different. For our observations with limited FOV ($\oslash = 44{^\circ}$) we calculated the corrected hourly rate $cHR$, what is the hourly rate corrected on the zenith distance of the radiant. It is called $corrected$ to be distinguished from visual zenith hourly rate (ZHR), although both are computed in the same way. Details can be found in \citet{Koten2020}. 

The activity profile recorded by the video cameras is shown in Figure~\ref{fig_gemchr}. The rates reached a maximum value of about 120 meteors per hour during the first night. The initial increase was slow as one of the stations started observation later due to the cloudy sky. A small dip after midnight was again caused by an interval of cloudy weather at the second site. The same is valid for the morning hours. Excluding a small dip around in the middle of the night and averaging the five highest values we got $cHR=115$. For comparison measurements by the IMO Video Meteor Network were added for the first night\footnote{https://www.imonet.org/reports/200612.html}. Although the absolute values are different, probably due to the different sensitivity of the cameras, there is good agreement with our profile. Both data sets show the peak around midnight. The IMO profile did not fall to zero in the morning as the network covers a broader range of the geographical longitudes.

The numbers of observed Geminid meteors per hour are significantly lower for the second night. It seems that the value of $cHR$ is slowly decreasing during the night. We adopt $cHR=40$ for this night.

\subsection{Candidates for pairs}

Based on the September $\epsilon$ Perseid observation \citep{Koten2017} when all the meteors occurred within 1.5 seconds, we chose 2-second separation as an initial threshold for the definition of the meteor pair. During 8 hours of the observations in the night of the maximum activity, we detected 18 such pairs. Just 7 of them were separated more than 1~second. 

This sample contains only meteors with a complete light curve. A whole luminous trajectory is recorded and the information about the photometric mass is complete. The time gap $\Delta t$ between two meteors was defined as the difference in the time of the first appearance of each of them. The minimal distance $\Delta D$ was measured as the distance between both meteoroids. If the first one disappeared before the beginning of the second one, its trajectory was prolonged, the position at the beginning of the second meteor estimated and the distance between both meteoroids computed. The typical duration of observed meteors was several tenths of a second.  The average time interval between them $\Delta t = 0.79$~s, average minimal distance $\Delta D = 61.0$~km. The list of all possible pairs with a time interval and spatial distance between two meteors, the photometric mass of the heavier meteoroid as well as the mass ratios is given in Table \ref{tab_pairs}. The mass ratio is defined as the photometric mass of the heavier meteoroid to the photometric mass of the lighter one within one pair.

\begin{table}
\caption{List of potential meteor pairs.}             
\label{tab_pairs}      
\centering          
\begin{tabular}{c c c c c c c c c} 
\hline

Meteor no.	&	Time			&	$\Delta$t		&	$\Delta$D	& $M_{H}$		& Mass 			\\
0C613xxx	&	[UT]			&	[s]				&	[km]		& [$10^{-3}$g]	& ratio			\\
\hline
040*, 041*	&	22:50:00.28		&	0.20			&	23.8		&	9.6			&	5.00		\\
069, 070	&	23:03:03.72		&	0.20			&	22.6		&	1.5			&	3.75		\\
090, 091	&	23:14:33.42		&	0.36			&	21.8		&	10.			&	2.08		\\
130, 131*	&	23:31:20.12		&	0.88			&	44.1		&	15.			&	2.14		\\
153*, 154	&	23:43:45.56		&	1.30			&	85.5		&	4.1			&	2.16		\\
196*, 197	&	0:01:21.12		&	0.14			&	77.8		&	9.			&	2.20		\\
255*, 256	&	0:28:46.68		&	1.08			&	137.9		&	65.			&	9.15		\\
283*, 284	&	0:39:01.48		&	1.64			&	117.7		&	5.2			&	4.35		\\
302A*, 302B* &	0:52:03.58		&	0.44			&	23.2		&	4.5			&	2.27		\\
308A, 308B	&	0:54:29.20		&	0.76			&	34.8		&	2.9			&	1.38		\\
350, 351	&	1:14:22.04		&	1.84			&	75.4		&	55.			&	5.00		\\
360A, 360B	&	1:21:27.72		&	0.48			&	53.3		&	570.		&	62.0		\\
361, 362*	&	1:21:46.94		&	1.50			&	88.6		&	4.9			&	1.69		\\
414, 415*	& 	1:45:27.76		&	0.76			&	80.0		&	7.1			&	1.03		\\
455, 456*	&	2:01:51.16		&	1.20			&	74.0		&	12.			&	2.22		\\
460, 461	&   2:03:52.58		&	0.78			&	53.5		&	19.			&	5.76		\\
505, 506*	&	2:25:33.32		&	0.20			&	41.8		&	6.2			&	3.10		\\
572, 573	&	3:13:24.92		&	1.56			&	68.8		&	21.			&	6.36		\\
631*, 632	&	4:16:08.62		&	0.30			&	64.1		&	98.			&	13.4		\\
\hline                  
260A*, 260B	&	0:30:21.83		&	0.60			&	38.5		&	5.6			&	2.66		\\
260B, 261*	&	0:30:22.43		&	0.30			&	91.0		&	8.7			&	1.56		\\
\hline                  
\end{tabular}
\tablefoot{The time marks the first appearance of the first meteor, $\Delta$t	is a time interval between both meteors, $\Delta$D is a minimal distance in 3D space. The error of $\Delta$D determination is up to 1~km. The mass ratio is defined as the ratio of bigger vs. smaller meteor photometric masses. The photometric mass of heavier meteoroids $M_{H}$ is provided too. Meteors marked with (*) are only single-station ones. The triplet of meteors observed around 0:32:22 is also listed.}
\end{table}

\begin{figure}
  \resizebox{\hsize}{!}{\includegraphics{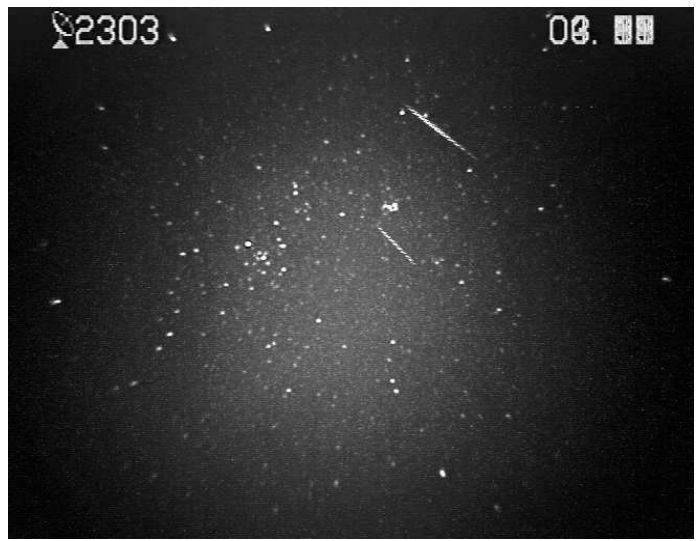}}
  \caption{The composite image of two Geminid meteors 06C13069 (in upper part of the image) and 06C13070 (in central part) representing one of the closest pairs observed in this study.}
  \label{fig_pair}
\end{figure}

\begin{figure*}
  \resizebox{\hsize}{!}{\includegraphics{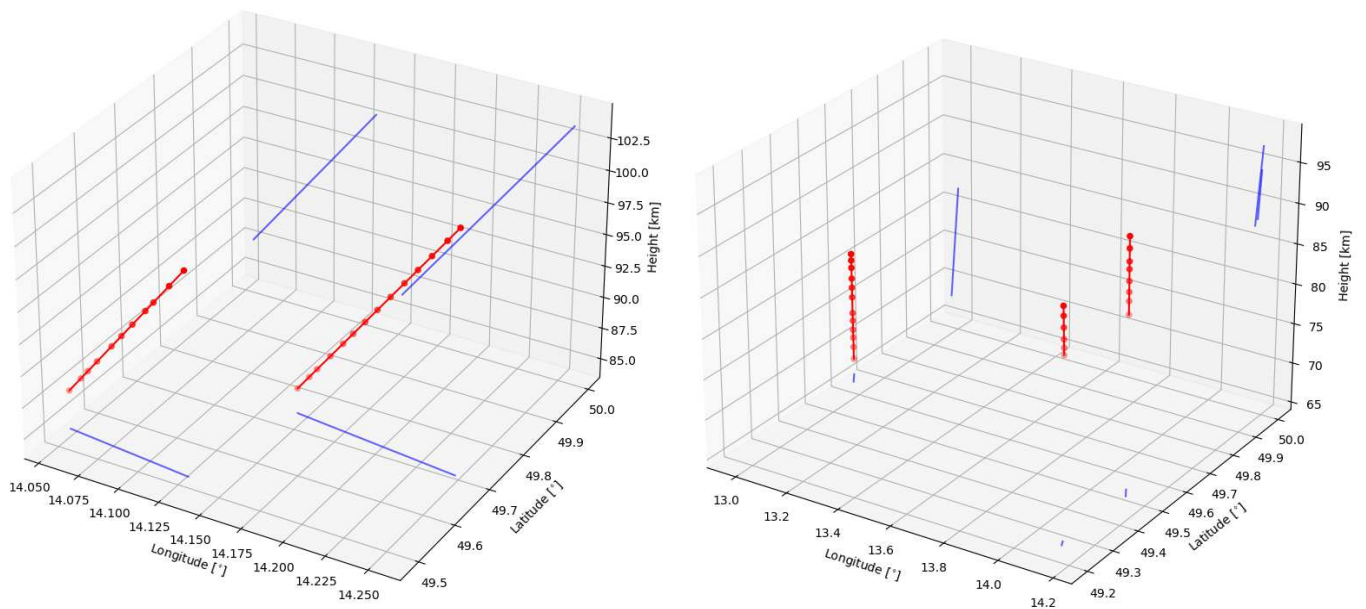}}
  \caption{The 3D map of the atmospheric trajectories (red dots) of close pair 069-070 (plot A with meteor 069 on the right and 070 on the left) and a triplet 06C13260A, 06C13260B and 06C13261 (plot B with meteors 260A on the right, 260B in the middle, 261 on the left) and their horizontal and vertical projections (blue lines). The x-axis represents the geographic longitude, the y-axis the geographic latitude, and the z-axis the height. Each red dot shows the position of the meteor at a given video frame. Two dots are separated by 0.04 s.}
  \label{fig_pairs3D}
\end{figure*}

Let us look at the meteors 06C13069 and 06C13070 as an example. Both meteors were recorded as double station cases. Firstly, meteor 069 appeared at 23:03:03.72~UT. Calculation shows that its beginning height was 105.1~km. Five frames later, i.e. $\Delta$t=0.2~s, became meteor 070 visible at height 101.6~km. As both meteors were seen simultaneously, we can directly determine their distance $\Delta$D= 22.6 km. Composite image shows that 069 was brighter (Figure \ref{fig_pair}). Its photometric mass was $2.1x10^{-3}$~g whereas for 070 we obtained value of $5.5x10^{-4}$~g (mass ratio = 3.75). Projected on the ground both trajectories are within a small area of 23.25 x 17.7 km, which is about 295 km$^{2}$. 3D view of the atmospheric trajectories of this pair is shown as plot A in Figure \ref{fig_pairs3D}. 

Similar calculations and visualizations were done for all pairs. Moreover, an unusual event was found occurring at 0:30:22~UT. Firstly, a single station meteor 06C13260A was recorded with a double station 06C13260B following 0.6~s later. The spatial distance between them was 38.5~km. After only 0,3~s another single station Geminid meteor 06C13261 was detected at the distance 91.0~km from the second meteor. The gap between the first and third meteors was 0.9~s and the distance 81.4~km. When projected on the ground, all three meteors cover an area of 4000 km$^{2}$. 3D view of all three meteors is shown as plot B in Figure~\ref{fig_pairs3D}. 

\section{Results}
\label{results}

Even if two meteors are observed simultaneously, it does not automatically mean that both form the physically connected pair. Their concurrent appearance can still be coincidental. This behaviour was analysed using statistical distributions in the past. Later, it was shown that such an approach can provide overestimated number of tied pairs. Therefore, we applied both approaches and provide a comparison of the results.

\subsection{Poisson distribution}

\cite{Porubcan1968}, and \cite{Porubcan2002} used several statistical methods to decide whether any pairing or grouping was seen among several meteor showers observed by radar or by TV cameras. Here, we apply this method to our video data. Note that the number of meteors in our sample is several times lower than the numbers available to \cite{Porubcan1968}. For example, he investigated six different samples of the Geminid shower containing between 800 and 2200 meteors. Another difference is that we concentrate only on very short gaps among meteors (1 and 2 seconds) as we are interested in very close pairs.

Traditionally, the number of observed meteors with a given time separation is compared with the expected Poisson distribution. The expected number of random appearances N is given by the equation \citep{Porubcan1968}:

\begin{equation}
N = n \frac{\mu^{x}}{x!} e^{-\mu}
\label{poisson}
\end{equation}

where $n$ is the number of intervals, $\mu$ is the number of meteors per interval and $x$ is the number of meteors within the cluster. For example, if the interval is 2 seconds, the number of intervals per hour is 1800. Having $cHR = 115$, $\mu$ = 0.064. Note that using $cHR$ means that number $N$ is also corrected to the zenith distance of the radiant.

The activity was not constant for the whole night as Figure~\ref{fig_gemchr} shows. Therefore, we calculated the expected number of pairs using Equation~\ref{poisson} for each time the $cHR$ is determined, summed all the cases, and compared with observed numbers. For example, for a 1-second interval between two consecutive meteors, we would expect $14.8\pm0.2$ random pairs during 8 hours long observation period. Results for three different intervals are shown in Table~\ref{tab_chr&pairs}. Note, that the observed numbers were also corrected on the zenith distance of the radiant. The excess of the number of actually observed pairs is seen in each of the selected intervals. If we look at the interval of 1 second, we can see that at least half of the observed pairs may be real, not random cases. On the other hand, the difference between the expected and observed number of pairs during the night following the maximum of activity is significantly smaller for all the intervals. 

\begin{table}
\caption{Expected and observed numbers of pairs.}             
\label{tab_chr&pairs}      
\centering          
\begin{tabular}{c | c c c} 
\hline
Night & 	Interval [s]	&	Expected		& 	Observed	\\
\hline
\multirow{4}{*}{13./14. Dec}
&	0.5				&	7.5$\pm$0.2		&	23			\\
&	1.0				&	14.8$\pm$0.4	&	33			\\
&	2.0				&	28.7$\pm$0.8	&	46			\\
\hline                  
\multirow{4}{*}{14./15. Dec}
&	0.5				&	2.8$\pm$0.2		&	6			\\
&	1.0				&	5.2$\pm$0.2		&	8			\\
&	2.0				&	10.2$\pm$0.6	&	14			\\
\hline
\end{tabular}
\tablefoot{Expected numbers of random pairs and observed numbers for different intervals between two consecutive Geminid meteors calculated according to the observed activity profile. Both, expected and observed numbers are corrected to the zenith distance of the radiant.}
\end{table}

Note, that according to Equation~\ref{poisson} the probability that three Geminid meteors are observed within 1 second is only 0.02 cases per hour, i.e. 0.16 cases per whole observing period. 

\subsection{Monte Carlo simulation}
\label{MCsimulation}

Although the results of this statistical test are promising, we are aware of the \citet{Sampson2007}'s paper, which warned that this method underestimates the number of random pairings and groupings. Therefore, we conducted additional tests.

The observed meteor pairs may be a result of chance. An alternative hypothesis is that (at least) some of them are physically related, i.e. they are produced by recent disruption of parent meteoroid. To assess these hypotheses, a Monte Carlo simulation was used. As a level of significance, we assumed the commonly used value of $5\%$. 100\,000 artificial random ``Geminid meteor showers'' which follow the activity profile of the observed 2006 Geminids shower were generated in the following way. 

We divided the observation time between $t_1=$~Dec.~13th,~21:18:53~UT and $t_2=$~Dec.~14th,~5:45:02~UT into 16 intervals (1898\,s each) and computed the number of meteors inside these intervals $N_i^{obs}$ with $i=1,\dots,16$. Next, random number $N_1\in \langle t_1,t_2\rangle$  with uniform distribution was generated and index $j$ of corresponding time interval was determined. Then another random number $N_2$ with uniform distribution was generated inside interval $\langle 0,\max\limits_{i=1\dots 16}(N_i^{obs})\rangle$.  If $N_2<N_j^{obs}$ then $N_1$ is accepted  as a time of artificial meteor, otherwise it is rejected. This was repeated until we obtain total number of 524 artificial meteor times. Finally, we counted occurrence of pairs in this set of 100\,000 artificial random Geminid meteor showers and derived corresponding probabilities.

We found that 81\,595 out of 100\,000 artificial random Geminid meteor showers contain at least one pair with $\Delta t\leq0.14$\,s. In other words, there is $\sim 82\%$ probability, that such a pair is a result of chance. For a pair with $\Delta t\leq0.2$\,s the corresponding probability is $\sim 91\%$. With the increasing value of $\Delta t$, the probability that the pair is a result of chance increases. In the case of a specific observed pair, it is therefore not possible to distinguish whether it was formed by fragmentation or only by chance. 

\begin{figure}
  \resizebox{\hsize}{!}{\includegraphics{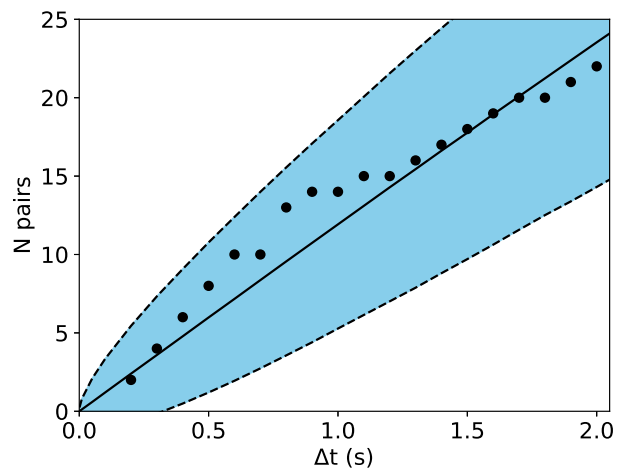}}
  \caption{Total number of pairs with the delay smaller than $\Delta t$. The solid line corresponds to the mean value from the Monte Carlo simulation and the blue area with dashed margins defines the $95\%$ interval of coincidence. Black dots represent the observed count of pairs. The hypothesis that the excess of observed pairs counts is a result of chance cannot be ruled out.}
  \label{Npairs_dt}
\end{figure}

The presence of physically related pairs would result in excess over the number of expected pairs for random distribution. We also tested this hypothesis. Figure \ref{Npairs_dt} shows a slight excess of observed pairs above expected counts according to Monte Carlo simulation. However, all the points are still inside the $95\%$ interval of coincidence. To be statistically significant they should lie outside the blue area. Therefore, again, we cannot rule out that all these pairs are only the result of chance. 

Let us focus on the triplet which was observed within 0.9\,s. The Monte Carlo simulation shows that at least one triplet with $\Delta t\leq 0.9$\,s occurs in $11\%$ of cases of artificial random Geminid meteor showers. This probability is lower than those of observed meteor pairs, but it is still higher than the assumed level of significance. Therefore, we cannot rule out that the triplet is only a result of chance.

\subsection{Geometry of pairs}

Despite the negative results of the Monte Carlo simulations, let us suppose that at least some of the observed pairs are real. If the fragmentation occurred in the interplanetary space days before the Earth's encounter, we may expect that lighter particles are affected more by the solar radiation pressure and pushed in the direction opposite to the Sun. Of course, if the ejection velocity of the lighter particle in the Sun direction is non-negligible, there should not be enough time to push this particle far away from the Sun than the heavier one. Note that the Poyinting-Robertson effect is negligible in this case. 

The acceleration depends on the particle's diameter $d$ and its density $\rho$ and can be estimated by equation \citep{Finson1968}:

\begin{equation}
a = \frac{3L_{\rm s}Q_{\rm pr}}{8\pi c r^2} \frac{1}{\rho d},
\label{eq_accel}
\end{equation}
where $L_{\rm  s}$ is the Solar luminosity, $Q_{\rm pr} = 1$ is the radiation pressure efficiency factor, $c$ is the speed of light, and $r$ is the heliocentric distance.

\begin{figure}
  \resizebox{\hsize}{!}{\includegraphics{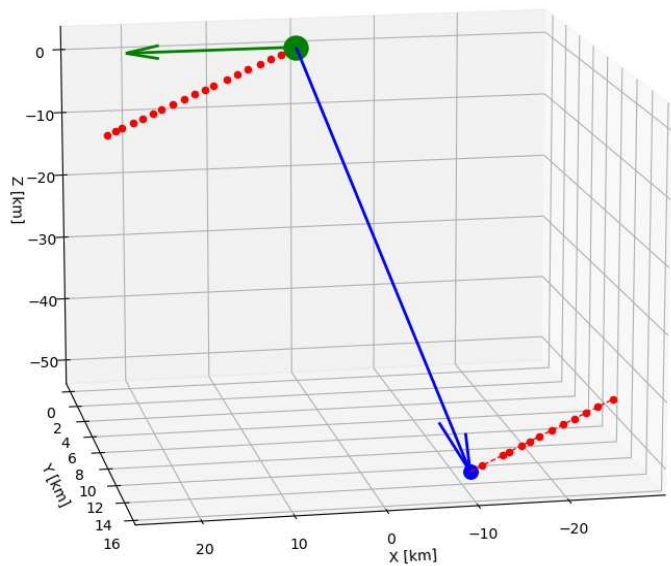}}
  \caption{This plot shows positions of the meteors (red dots) belonging to the pair candidate 360A (lower right), 360B (upper left) in a rectangular coordinated system with an origin connected with the first position of the heavier particle (360B) which is shown as a green circle. A smaller blue circle shows the theoretical position of the particle 360A at the same time extrapolated according to its motion. The blue vector illustrates the relative position of particle 360A against 360B. The x-axis is oriented into the direction of the Sun as the green vector shows. The vectors represent an angle between lighter particle, heavier particle, and the Sun.}
  \label{trajectoryXYZ}
\end{figure}

The position of each meteor point is given by its geographic longitude, latitude, and height above the ground. A rectangular coordinate system (X, Y, Z) was defined for each pair with an origin at the first point of a heavier particle. Azimuth and elevation of the Sun from this point were calculated. The original coordinates were transformed into the new system with a direction vector to the Sun to be the same as the X-axis. The position of the lighter particle at the same time was calculated according to its movement. An angle $\theta$ between the direction to the Sun and the direction to the lighter particle is visible in this representation. Figure~\ref{trajectoryXYZ} shows the case of the pair 360A, 360B. 

The value $\theta = 100.7^{\circ}$ was found for the pair 360A, 360B. It means that the heavier particle is closer to the Sun than the lighter one. The mass ratio is 62.0. The absolute distance between particles $\Delta D \sim$53~km, the radial component is $D_{R} \sim$ 10~km. Using Equation~\ref{eq_accel} and assuming a density of Geminids to be 2600~kg.m$^{-3}$ \citep{Borovicka2010}, the acceleration of i-th particle can be computed from $a_{i} = \frac{2.6x10^{-9}}{d_{i}}$, where $d_{i}$ is the particle diameter. The radial distance of 10~km is reached within 38.5~hours in the case of zero ejection velocity. Because the velocity of Geminid meteoroids is about 35 $km.s^{-1}$, the fragmentation can occur at a distance of 4.9 million km from the Earth. As the zero ejection velocity is improbable, we set two extreme limits on the determination of the ejection time. Let suppose the ejection velocity to be $\pm 1~m.s^{-1}$. That such assumption is realistic, is shown for example in the paper of \citet{Hapgood1981}, which supposed velocities between 0.6 and 6.6 $m.s^{-1}$. A negative value means ejection in the direction to the Sun, positive in the opposite direction. Depending on the direction, the observed radial distance is reached in the time range between 3 hours and 22 days.

\begin{figure}
  \resizebox{\hsize}{!}{\includegraphics{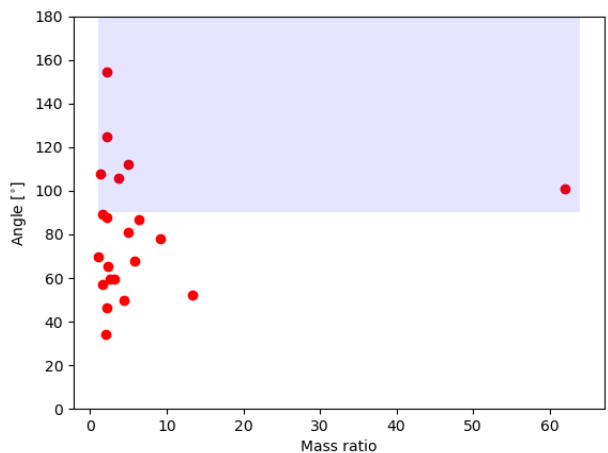}}
  \caption{The mass ratio and the angle between fragments and the Sun. Light blue rectangle represent an area, where the fragments separated from the same original body would be expected.}
  \label{angles}
\end{figure}

We determined the mutual positions of potential pairs and calculated $\theta$ angle for all the candidates. The results are summarized in Figure~\ref{angles}. There are two segments in this plot. The pairs having heavier member closer to the Sun than the lighter one can be found in the area represented by the blue rectangle. Only 6 of 21 candidates (the triplet is counted as three pairs in this analysis) are plotted within this segment. Their geometrical properties, time from possible fragmentation, and its distance are listed in Table~\ref{tab_EarthDist}. The majority of the candidates have $\theta < 90^{\circ}$, although few of them are very close to the threshold. More than half of the cases represent a situation when the less mass particle is closer to the Sun. 

\begin{table*}
\caption{Properties of candidates with $\theta > 90^{\circ}$.}             
\label{tab_EarthDist}      
\centering          
\begin{tabular}{c c c c c c c c} 
\hline

Meteor no.	&	Mass ratio	&	Angle $\theta$	&	Radial distance		&	Time$_{0}$	&	Time$_{-1}$		& 	Time$_{+1}$	&	Distance from Earth			\\
0C613xxx	&				&	[${\circ}$]		&	[km]				&	[hour]		&	[hour]			&	[hour]		&	[million km]				\\
\hline
069, 070	&	3.75		&	105.7			&	6.1					&	37.9		&	1.7				&	422			&	4.8							\\
153, 154	&	2.16		&	124.7			&	48.7				&	118.8		&	13.4			&	1042		&	15.0						\\
308A, 308B	&	1.38		&	107.8			&	10.6				&	123.2		&	2.9				&	5000		&	15.5						\\
350, 351	&	5.0			&	112.1			&	28.4				&	89.7		&	7.9				&	1015		&	11.3						\\
360A, 360B	&	62.0		&	100.7			&	10					&	38.5		&	2.8				&	531			&	4.9							\\
455, 456	&	2.22		&	154.6			&	67.2				&	163.4		&	18.4			&	1442		&	20.6						\\
\hline                  
\end{tabular}
\tablefoot{Columns $Time$ show the times necessary to reach radial distance between both particles -- $Time_{0}$ in case of zero ejection velocity, $Time_{-}$, $Time_{+}$ extreme cases of $1~m.s^{-1}$ ejection velocities, '+' in the Sun direction, '-' in opposite direction.}
\end{table*}

To understand what this result means we extended a Monte Carlo test which was done in Section~\ref{MCsimulation}. The times of the meteors were generated in the same way. For each meteor also a mass was generated randomly from the range of the observed masses. Moreover, the position of each meteor was also randomly determined inside the cube of a size of 80 km. The size of the cube was determined as the diameter of the field-of-view of the video camera at the distance of the observed meteors. Then the pairs with a time gap shorter than 2~seconds were search. For such a pair, the relative position between them was determined. Finally, the angle Sun - heavier particle - lighter particle was determined. This simulation was run 100 000 times.

\begin{figure}
  \resizebox{\hsize}{!}{\includegraphics{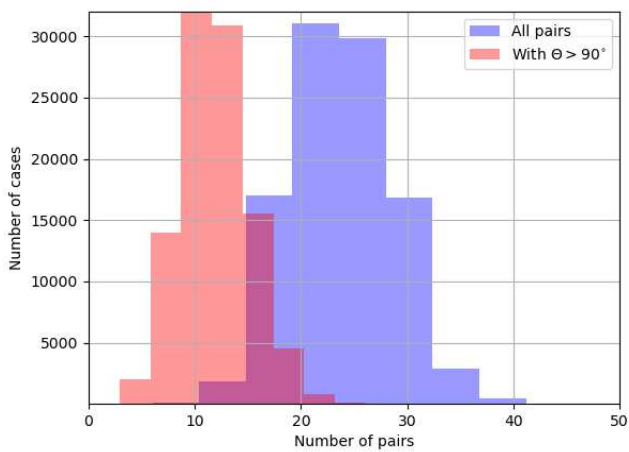}}
  \caption{Random numbers of the pairs and pairs with $\theta>90^{\circ}$ among 524 observed meteors.}
  \label{hist_angles}
\end{figure}

The results of the simulation are shown in Figure~\ref{hist_angles}. For each run, we found 23.6 pairs within 2 seconds. The observed number of 21 pairs fitted well inside the 2-$\sigma$ interval ($\sigma=4.7$). Again, we cannot exclude that all the observed pairs were actually random appearances of physically unconnected meteors. The simulation shows that in 11.8 of such cases, i.e. 50\% of them, the lighter particle is at a greater distance than the heavier one. The observed number of such cases is again inside the 2-$\sigma$ interval ($\sigma=3.4$). Therefore, the arrangement of the particles inside the pair tells us nothing about their origin -- besides the unknown ejection velocity during hypothetical fragmentation event -- we cannot rule out their coincidental origin.

\section{Conclusions}
\label{conclusions}

The visual inspection of the 2006 Geminid archive data during its digitalization shows a higher than a usual number of meteor pairs appearing within less than 2 seconds. It looked like a promising sample of interesting data for the study of the meteor pairing. The traditional way of the analyses using Poisson statistics supported this suspicion. It showed that there is a significant excess of the pairs in comparison with the randomly expected number of the cases. 

The atmospheric trajectories computation shows that the potential pairs were very close not only temporally but also spatially. The closest pairs were separated only by few tens of kilometers. The distances were comparable or even smaller than separations among the fragments of well documented September epsilon Perseid meteor cluster which was proven as a real case of the meteor grouping \citep{Koten2017}.

However, the Monte Carlo simulations made on the sample of 100 000 artificial Geminid showers with similar activity profiles showed that even the most promising cases may be still results of random apparency of two meteoroids without any physical connection. Comparison of the observed and expected number of pairs showed small excess of the observed cases but it was still inside of $95\%$ interval of coincidence. Therefore, we cannot rule out that all the cases are random appearances. The results of this study confirmed the suspicion of \citet{Sampson2007} that the Poisson statistics using fixed intervals cannot be used for this kind of analyses as it underestimates the numbers of randomly observed pairs.

If the fragmentation of the meteoroid in the interplanetary space occurred before the encounter with Earth we would expect the smaller meteoroid to be pushed more in the anti-Sun direction than the bigger one. Such behaviour was observed e.g. in the case of September epsilon Perseid cluster \citep{Koten2017}. An extension of the Monte Carlo simulation shows that the observed data are in agreement with the randomly generated population of Geminid meteors according to the meteor shower activity. Therefore, it is not possible to prove that the observed pair is a real pair or it represents a coincidental appearance of two independent meteors. 

It was mentioned above that \cite{Porubcan2002} identified non-random groups of the meteoroids within the core of young streams. Those streams were 1969 and 1999 Leonids which were caused by 1 and 3 revolutions old filaments \citep{McNaught1999}. As the Geminid meteor stream is a few thousand years old, it is not young from this point of view. It may be a similar case as the Lyrid shower in Porub\v{c}an' paper for which also no evidence for groupings was found. Moreover, as \cite{Plavec1953} already notes, the dispersion of the pairs is relatively fast. The only exceptions are pairs with components of equal masses. He compared several meteor showers and found that the relative rate of disintegration is about eight times higher for Lyrids or Geminids in comparison with Leonids. Finally, the strength of Geminid meteoroids may be another reason for the low abundance of the pairs within the stream. Due to high strength, there could not be a sufficient number of fragmentation events in the vicinity of the Earth to create a detectable number of the pairs.

As a lesson from this work, there are few recommendations for future observation campaigns which can increase the probability of successful detection of the pairs and groups. The wider field-of-view (FOV) cameras can be useful. It allows detection of a higher number of the meteors among them the pairs may be present. Even more spatially distant meteors can be still considered as pairs and those can be missed using very narrow FOV cameras. On the other hand, using very wide FOV and even all-sky cameras usually can eliminate a lot of fainter meteors. Also, the selection of the target meteor shower can improve chances for pairs and groups detection. Using a relative rate of disintegration defined by \cite{Plavec1953} shows that Perseid, Taurids, Leonids, or Draconids are better candidates than Geminids.

Altogether, we conclude that we did not find firm evidence for the existence of the pairs and groups within the Geminid meteor shower, at least its 2006 appearance.

\begin{acknowledgements}
This work was supported by the Grant Agency of the Czech Republic grants 20-10907S and the institutional project RVO:67985815. We thank to an anonymous referee, who helped us to improve the paper. 
\end{acknowledgements}

\bibliographystyle{aa}
\bibliography{00pkbib}

\end{document}